\documentclass[a4paper, 12pt]{article}

\usepackage{tikz}
\usetikzlibrary{arrows,decorations.markings}
\usepackage{latexsym,amsmath,amsfonts,amssymb}
\usepackage[latin1]{inputenc}
\usepackage[american]{babel}
\usepackage{bbm}
\usepackage{hyperref}
\usepackage{mathrsfs}

\newcommand{\dvol}{d\mathrm{vol}}

\newcommand{\Vol}{\mathrm{Vol}}

\newcommand{\parfrac}[2]{\frac{\partial #1}{\partial #2}}

\newcommand{\du}[2]{_{#1}^{\phantom{#1}#2}}

\newcommand{\wt}{\widetilde}
\newcommand{\wh}{\widehat}
\newcommand{\wb}{\overline}
\newcommand{\matht}[1]{\ensuremath{\boldsymbol{#1}}}

\newcommand{\eg}{\textit{e.g.}}

\numberwithin{equation}{section}

\newcommand{\mat}[1]{\begin{pmatrix} #1 \end{pmatrix}}
\newcommand{\smat}[1]{\big( \begin{smallmatrix} #1 \end{smallmatrix} \big)}
\newcommand{\be}{\begin{equation}} \newcommand{\ee}{\end{equation}}
\newcommand{\bea}{\begin{equation} \begin{aligned}} \newcommand{\eea}{\end{aligned} \end{equation}}

\newcommand{\cA}{\mathcal{A}}
\newcommand{\cB}{\mathcal{B}}

\newcommand{\cD}{\mathcal{D}}

\newcommand{\cF}{\mathcal{F}}
\newcommand{\cG}{\mathcal{G}}

\newcommand{\cI}{\mathcal{I}}

\newcommand{\cK}{\mathcal{K}}
\newcommand{\cL}{\mathcal{L}}
\newcommand{\cM}{\mathcal{M}}
\newcommand{\cN}{\mathcal{N}}

\newcommand{\cP}{\mathcal{P}}
\newcommand{\cQ}{\mathcal{Q}}

\newcommand{\cS}{\mathcal{S}}

\newcommand{\cV}{\mathcal{V}}

\newcommand{\cZ}{\mathcal{Z}}

\newcommand{\bB}{\mathbb{B}}
\newcommand{\bC}{\mathbb{C}}

\newcommand{\bH}{\mathbb{H}}

\newcommand{\bP}{\mathbb{P}}

\newcommand{\bR}{\mathbb{R}}

\newcommand{\bZ}{\mathbb{Z}}

\newcommand{\fg}{\mathfrak{g}}

\newcommand{\fm}{\mathfrak{m}}
\newcommand{\fn}{\mathfrak{n}}

\newcommand{\fq}{\mathfrak{q}}

\newcommand{\sT}{{\sf{T}}}
\newcommand{\unit}{\mathbbm{1}}

\newcommand{\fsu}{\mathfrak{su}}

\def\repa{\raise4pt\hbox{$\square$}\mkern-14mu\raise-4pt\hbox{$\square$}}
\def\repab{\overline{\raise4pt\hbox{$\square$}\mkern-14mu\raise-4pt\hbox{$\square$}\mkern-1mu}}

\def\smileface{\ensuremath{\hbox{\large$\bigcirc$}\mkern-15mu\raise-1pt\hbox{\scriptsize$\smallsmile$}%
\mkern-10mu\raise4pt\hbox{..}\mkern4mu}}
\def\frownface{\ensuremath{\hbox{\large$\bigcirc$}\mkern-15mu\raise-1pt\hbox{\scriptsize$\smallfrown$}%
\mkern-10mu\raise4pt\hbox{..}\mkern4mu}}

\DeclareMathOperator{\Tr}{Tr}

\DeclareMathOperator{\re}{\mathbb{R}e}
\DeclareMathOperator{\im}{\mathbb{I}m}
\DeclareMathOperator{\Li}{Li}

\begin{document}

\thispagestyle{empty}
\begin{flushright}
SISSA  34/2017/FISI
\end{flushright}
\vspace{10mm}
\begin{center}
{\huge  Black Hole Entropy in Massive Type IIA} 
\\[15mm]
{Francesco Benini$^{1,2,3}$, Hrachya Khachatryan$^2$, Paolo Milan$^2$}
\vskip 6mm
 
\bigskip
{\it
$^1$ Institute for Advanced Study, Princeton, NJ 08540, USA \\[.5em]
$^2$ SISSA, via Bonomea 265, 34136 Trieste, Italy \\[.5em]
$^3$ INFN, Sezione di Trieste, via Valerio 2, 34127 Trieste, Italy \\
}
\vskip 6 mm

\bigskip
\bigskip

{\bf Abstract}\\[5mm]
{\parbox{14cm}{\hspace{5mm}

We study the entropy of static dyonic BPS black holes in AdS$_4$ in 4d $\mathcal{N}=2$ gauged supergravities with vector and hyper multiplets, and how the entropy can be reproduced with a microscopic counting of states in the AdS/CFT dual field theory. We focus on the particular example of BPS black holes in AdS$_4 \times S^6$ in massive Type IIA, whose dual three-dimensional boundary description is known and simple. To count the states in field theory we employ a supersymmetric topologically twisted index, which can be computed exactly with localization techniques. We find perfect match at leading order.

}
}
\end{center}
\newpage
\pagenumbering{arabic}
\setcounter{page}{1}
\setcounter{footnote}{0}
\renewcommand{\thefootnote}{\arabic{footnote}}

\tableofcontents

\section{Introduction}

One of the important testing grounds for string theory, as a theory of quantum gravity, is the physics of black holes. In string theory they can be formed from systems of branes which, in turn, admit a description in terms of a worldvolume gauge theory. This provides us with a powerful alternative point of view, besides the gravitational description, much more amenable to a quantum treatment. In fact, this framework was first used by Strominger and Vafa \cite{Strominger:1996sh} to show that, within string theory, one can give a microscopic statistical interpretation to the thermodynamic Bekenstein-Hawking entropy \cite{Bekenstein:1973ur, Hawking:1974sw} of BPS black holes in flat space. This is possible because string theory embeds gravity into a consistent quantum theory.

Many different setups have been analyzed since then, including quantum corrections, to an impressive precision \cite{Dijkgraaf:1996it, Shih:2005qf, Pioline:2005vi, David:2006yn, Sen:2008ta} (see \eg{} \cite{Sen:2007qy} for more references). In essentially all examples, the microscopic counting is performed in a 2d CFT that appears---close to the black hole horizon---as one plays with the moduli available in string theory (and takes advantage of various dualities). In fact, also the entropy of BPS black holes in AdS$_3$ is well understood, since the microscopic state counting can be performed in the 2d CFT related to AdS$_3$ by the holographic (AdS/CFT) duality \cite{Maldacena:1997re}.

For black holes in AdS in four and more dimensions the situation is different, because in general there is no regime in which the black holes are described by a 2d CFT. On the other hand, AdS/CFT provides a non-perturbative definition of the entire quantum gravity in AdS in terms of a standard QFT living at its boundary.%
\footnote{By contrast, for black holes in flat space one uses a different QFT description for each black hole.}
Therefore one would expect the black holes to appear as ensembles of states with exponential degeneracy in the QFT.

The first successful entropy match in AdS$_4$ has been done in \cite{Benini:2015eyy} (see \cite{Hosseini:2016tor, Hosseini:2016ume, Benini:2016hjo, Benini:2016rke, Hosseini:2016cyf, Cabo-Bizet:2017jsl, Nian:2017hac} for subsequent work and generalizations). The match is for static dyonic BPS black holes in AdS$_4 \times S^7$ in M-theory. They can be more conveniently described by a consistent truncation to a 4d $\cN=2$ gauged supergravity with three Abelian vector multiplets---the so-called STU model \cite{Duff:1999gh}. The microscopic counting is performed in the dual 3d boundary theory, the ABJM theory \cite{Aharony:2008ug}. The black hole microstates are identified with ground states of the 3d QFT placed on $S^2$ (or a Riemann surface, depending on the horizon topology) and topologically twisted. Such states can be conveniently counted by an index, defined in \cite{Benini:2015noa} and called ``topologically twisted index'':
\be
Z(\fn_a, \Delta_a) = \Tr\, (-1)^F\, e^{-\beta H} \, e^{i \Delta_a \fq_a} \;.
\ee
Here $\Delta_a$ are chemical potentials for the electric charges $\fq_a$, while $H$ is the Hamiltonian on $S^2$ which depends on the integers $\fn_a$ associated to the magnetic charges of the black hole. Because of supersymmetry, only the states with $H=0$ contribute. In order to make contact with weakly-curved gravity, one should take a large $N$ limit in the QFT. Assuming that at leading order there are no dangerous cancelations due to $(-1)^F$,%
\footnote{An argument, similar to the one in \cite{Sen:2009vz}, was given in \cite{Benini:2015eyy} that $(-1)^F = 1$ on states of the single-center black hole, while one expects $(-1)^F$ to be $\pm1$ on states related to multi-center black holes and hair.}
the quantum degeneracies can be extracted with a Fourier transform, which at large $N$ becomes a Legendre transform:
\be
S_\text{BH} = \log Z(\fn_a, \wh \Delta_a) - i \sum\nolimits_a \wh \Delta_a \fq_a
\ee
with $\wh\Delta_a$ such that the right-hand-side is extremized. For black holes in AdS$_4 \times S^7$ this computation exactly reproduces the Bekenstein-Hawking entropy.

A similar match for BPS black holes with non-compact hyperbolic horizon has been performed in \cite{Cabo-Bizet:2017jsl}. An attempt to match some subleading corrections (scaling as $\log N$) has been made in \cite{Liu:2017vll, Jeon:2017aif}. An interesting observation about the case of black holes in AdS$_5$ has been put forward in \cite{Hosseini:2017mds}.

No many other examples have been checked so far.%
\footnote{See the recent work \cite{Azzurli:2017kxo} for more examples.}
However, putting together the results in \cite{DallAgata:2010ejj, Benini:2015eyy, Hosseini:2016tor, Benini:2016rke} one expects the match to work for static BPS black holes that can be described in consistent truncations to 4d $\cN=2$ gauged supergravities with only vector multiplets. In the gravity description, the near-horizon region to the BPS black holes is controlled by attractor equations \cite{DallAgata:2010ejj}. Schematically (and in a frame with purely electric gauging) the Bekenstein-Hawking entropy is proportional to the horizon area, and is given by the value of the function
\be
\cS \; \propto\;  i \sum\nolimits_a \frac{\fn_a \partial_a \cF(X) - \fq_a X^a}{\sum_b X^b}
\ee
at its critical point. Here $(\fn_a, \fq_a)$ are the magnetic and electric charges of the black hole (in suitable units), $X^a$ are the sections parametrizing the scalars in vector multiplets, and $\cF$ is the prepotential. One identifies $\Delta_a = X^a/ \sum_b X^b$, and using that the prepotential is homogeneous of degree two one can write
\be
\cS \;\propto\; \sum\nolimits_a \Big( i \, \fn_a \frac{\partial\cF(\Delta)}{\partial \Delta_a} - i \, \fq_a \Delta_a \Big) \;.
\ee
On the other hand it has been shown in \cite{Hosseini:2016tor} that, for a large class of quiver gauge theories appearing in AdS/CFT pairs, the large $N$ limit of the index is related to the large $N$ limit of the $S^3$ free energy $F_{S^3}$ by
\be
\log Z = \frac\pi2 \sum\nolimits_a \fn_a \parfrac{F_{S^3}}{\Delta_a} \;.
\ee
Thus, provided one verifies the proportionality between the supergravity prepotential and the $S^3$ free energy---which is a property of the conformal vacuum and has nothing to do with black holes---one has also matched the entropy of static dyonic BPS black holes.

In general, however, consistent truncations contain also hypermultiplets. They can give mass to some of the vector multiplets and affect the values of the vector multiplet scalars at the horizon, hence the simple argument presented above does not go through. The purpose of this paper is to investigate whether the entropy of AdS$_4$ black holes can be microscopically reproduced also in such more general theories. We consider a particularly interesting example: black holes in AdS$_4\times S^6$ in massive Type IIA. The AdS$_4$ vacuum has been recently constructed in \cite{Guarino:2015jca} and the dual three-dimensional SCFT has been identified as well. It is a 3d $\cN=2$ $SU(N)_k$ Chern-Simons gauge theory with three adjoint chiral multiplets and superpotential $W = \Tr X[Y,Z]$ (see Figure \ref{fig: quiver}). Besides, the near horizon geometries of static dyonic BPS black holes have been identified in \cite{Guarino:2017pkw} (see also \cite{Guarino:2017eag}). Making use of the attractor equations with hypermultiplets and dyonic gaugings \cite{Klemm:2016wng}, we are able to reproduce---at leading order---the entropy of those black holes from a microscopic counting.

We also stress that the black holes considered here are in massive Type IIA \cite{Romans:1985tz}, as opposed to M-theory. As a result, the entropy scales as $N^{5/3}$ as opposed to $N^{3/2}$. Yet, the microstate counting works perfectly and this gives us confidence on the robustness of the proposal in \cite{Benini:2015eyy}.

The paper is organized as follows. In Section \ref{sec: SUGRA} we describe the near-horizon geometries of static dyonic BPS black holes in AdS$_4 \times S^6$. We recast their entropy in the form of the solution to an extremization problem. In Section \ref{sec: FT} we compute the index in the field theory, at leading order in $N$, and express again the microstate degeneracy as the solution to an extremization problem. In Section \ref{sec: comparison} we show that the two problems coincide. We conclude in Section \ref{sec: conclusions}.

\vspace{1em}
\noindent
{\it Note added.} When this work was under completion, we became aware of the related works \cite{Azzurli:2017kxo} and \cite{Hosseini:2017fjo} that overlap with ours. We have coordinated the release of our work with \cite{Hosseini:2017fjo}.


\section{Dyonic black holes in massive Type IIA}
\label{sec: SUGRA}

We study BPS black holes in massive Type IIA on AdS$_4 \times S^6$. The supersymmetric AdS$_4$ vacuum, corresponding to the near-horizon geometry of $N$ D2-branes in the presence of $k$ units of RR 0-form flux (the Romans mass \cite{Romans:1985tz}), has been constructed in \cite{Guarino:2015jca}. The $S^6$ is squashed, as a squashed $S^2$ bundle over $\bC\bP^2$, and it preserves $U(1)_R \times SU(3)$ isometry. The first factor is an R-symmetry, and the solution preserves $4+4$ supercharges.

We are interested in static dyonic BPS black holes in this geometry, and they are more conveniently described within a consistent truncation to 4d $\cN=2$ gauged supergravity. In particular, massive Type IIA on AdS$_4 \times S^6$ admits a consistent truncation to $ISO(7)$-dyonically-gauged 4d $\cN=8$ supergravity \cite{Guarino:2015jca}, where $ISO(7) = SO(7) \ltimes \bR^7$ (see also \cite{DallAgata:2014tph, Guarino:2015qaa}). This theory has three AdS solutions, and the one we are interested in preserves $\cN=2$ supersymmetry and a $U(1)_R \times SU(3)$ subgroup of $ISO(7)$. Dyonic black holes generically break $U(1)_R \times SU(3)$ to its maximal torus, and can be described by a further consistent truncation to a 4d $\cN=2$ gauged supergravity with vector and hyper multiplets. Such truncations are characterized by a subgroup $G_0 \subset ISO(7)$ under which all fields are neutral. We are thus interested in the case where $G_0 = U(1)^2$. Such an $\cN=2$ truncation contains three vector multiplets and one hypermultiplet, and what is gauged is a group $\bR\times U(1)$ of isometries of the hypermultiplet moduli space \cite{Guarino:2017pkw}.

When dealing with 4d $\cN=2$ supergravity, it is convenient to use the language of special geometry \cite{deWit:1984wbb, Strominger:1990pd, Andrianopoli:1996cm}.%
\footnote{We follow the notation of \cite{Klemm:2016wng}.} Let us restrict to the case with Abelian gauge fields, then the formalism is covariant with respect to symplectic $Sp(2n_V+2,\bZ)$ electric-magnetic transformations ($n_V$ is the number of vector multiplets). We use a notation $V^M = (V^\Lambda, V_\Lambda)$ for symplectic vectors, where $\Lambda = 0, \dots, n_V$, and define the symplectic scalar product
\be
\langle V, W \rangle = V^M \Omega_{MN} V^N = V_\Lambda W^\Lambda - V^\Lambda W_\Lambda
\ee
in terms of the symplectic form $\Omega = \smat{ 0 & -\unit \\ \unit & 0}$.

First, the complex scalars $z^a$ in vector multiplets (with $a=1,\dots, n_V$) describe a special K\"ahler manifold $\cM_\text{SK}$. We can give it a (redundant) parametrization in terms of holomorphic sections $X^\Lambda$. The holomorphic sections are collected into a covariantly-holomorphic symplectic vector
\be
\cV = e^{\cK(z^a, \bar z^a)/2} \mat{ X^\Lambda(z^a) \\ \cF_\Lambda(z^a) }
\ee
with $D_{\bar a} \cV = \partial_{\bar a} \cV - \frac12 (\partial_{\bar a} \cK)\cV = 0$.
Here $\cK(z^a, \bar z^a) = -\log \big[ i ( \cF_\Lambda \wb X^\Lambda - X^\Lambda \wb \cF_\Lambda) \big]$ is the K\"ahler potential for the metric on $\cM_\text{SK}$, namely $ds_\text{SK}^2 = - ( \partial_a \partial_{\bar b} \cK)  dz^a d\bar z^{\bar b}$, while $\cF_\Lambda = \partial_\Lambda \cF$ are the derivatives of the prepotential $\cF$. Thus the covariantly-holomorphic sections satisfy $\langle \cV, \wb\cV \rangle = -i$. In addition to $z^a$, the vector multiplets contain gauge fields $A^a$ which, together with the graviphoton $A^0$, form a symplectic vector $\cA^M = (A^\Lambda, \wt A_\Lambda)$ where $\wt A_\Lambda$ are dual to $A^\Lambda$ under electric-magnetic duality.

In our case%
\footnote{More details about this gauged supergravity and its action can be found in \cite{Guarino:2017pkw}.}
$n_V=3$ and the special K\"ahler manifold is $\cM_\text{SK} = \big( SU(1,1)/U(1) \big)^3$ parametrized by $\{z^a\}_{a=1,2,3}$. The prepotential is
\be
\cF = -2 \sqrt{ X^0 X^1 X^2 X^3 }
\ee
(as in the STU model \cite{Duff:1999gh}) and the holomorphic sections can be parametrized as
\be
X^\Lambda = \big( - z^1 z^2 z^3,\, -z^1,\, -z^2,\, -z^3 \big) \;,\qquad\qquad \cF_\Lambda = \big( 1,\, z^2z^3,\, z^1z^3,\, z^1z^2 \big) \;.
\ee
In other words $X^1X^2X^3/X^0 = 1$. The K\"ahler potential is $\cK = - \sum_{a=1}^3 \log\big( 2 \im z^a \big)$ and the metric is
\be
ds^2_\text{SK} = \frac14 \sum_{a=1}^3 \frac{dz^a\, d\bar z^{\bar a}}{(\im z^a)^2} \;.
\ee
Thus the scalars $z^a$ live on the upper half plane.

Second, the real scalars $q^u$ in hypermultiplets (with $u=1,\dots, 4n_H$ and $n_H$ is the number of hypermultiplets) describe a quaternionic K\"ahler manifold $\cM_\text{QK}$. The dyonic gauging involves an isometry of $\cM_\text{QK}$ with associated commuting Killing vectors $k_\alpha$ (where $\alpha$ parametrizes the isometry generators). The specific gauging is described by an embedding tensor $\Theta\du{M}{\alpha}$ that contains information about the coupling of gravitini and hypermultiplets to the gauge fields. One requires the locality constraint $\langle \Theta^\alpha, \Theta^\beta\rangle=0$ that ensures the existence of a frame where the gauging is purely electric \cite{deWit:2005ub}. Hence, one constructs a symplectic Killing vector $\cK_M^u = \Theta\du{M}{\alpha} k^u_\alpha$ and then the covariant derivatives of the scalars $q^u$ are given by\be
Dq^u = dq^u - \langle \cA, \cK^u\rangle = dq^u + A^\Lambda \Theta\du{\Lambda}{\alpha} k_\alpha^u - \wt A_\Lambda \Theta^{\Lambda\alpha} k_\alpha^u \;.
\ee
The isometries of $\cM_\text{QK}$ descend from $SU(2)$-triplets $P^x_\alpha$ of moment maps, where $SU(2)$ acts on the supercharges and $x=1,2,3$. Once again, one can use the embedding tensor to construct a symplectic vector
\be
\cP^x_M = \Theta\du{M}{\alpha} P^x_\alpha \;.
\ee
The $SU(2)$ index $x$ is related to an $SU(2)$ bundle over $\cM_\text{QK}$, and one can thus perform local $SU(2)$ rotations.

In our case $n_H=1$ and the hypermultiplet manifold is $\cM_\text{QK} = SU(2,1) / \big( SU(2) \times U(1) \big)$. We parametrize it with $q^u = (\sigma, \phi, \zeta, \tilde\zeta)$ and its metric is given by
\be
ds^2_\text{QK} = h_{uv} dq^u dq^v = \frac14\, e^{4\phi} \Big( d\sigma + \tfrac12 \big(\zeta d\tilde\zeta - \tilde\zeta d\zeta \big) \Big)^2 + d\phi^2 + \frac14\, e^{2\phi} \big( d\zeta^2 + d\tilde\zeta^2 \big) \;.
\ee
The dyonic gauging involves an Abelian $\bR \times U(1)$ isometry of $\cM_\text{QK}$ with Killing vectors
\be
k_\bR = \partial_\sigma \;,\qquad\qquad k_{U(1)} = \zeta \partial_{\tilde\zeta} - \tilde\zeta \partial_\zeta \;.
\ee
Here $\alpha = \bR, U(1)$. They descend from moment maps
\bea
\label{moment maps}
P_\bR^+ &= 0 \;,\qquad\qquad & P_{U(1)}^+ &= e^\phi (\tilde\zeta - i \zeta) \;, \\
P_\bR^3 &= - \frac12 e^{2\phi} \;,\qquad\qquad & P_{U(1)}^3 &= 1 - \frac14 e^{2\phi} (\zeta^2 + \tilde\zeta^2) \;,
\eea
where $P_\alpha^+ = P_\alpha^1 + i P_\alpha^2$. The embedding tensor is
\be
\Theta^{M\alpha} = \mat{ \Theta^{\Lambda\alpha} \\ \hline \Theta\du{\Lambda}{\alpha}} = \left( \begin{array}{cccc|cccc} m & 0 & 0 & 0 & g & 0 & 0 & 0 \\ 0 & 0 & 0 & 0 & 0 & g & g & g \end{array} \right)^\sT
\ee
where $g,m$ are the electric and magnetic coupling constants, respectively, with dimension of mass. We will assume $g,m>0$. Notice that the hypermultiplet is charged only under one linear combination of the three $U(1)$ gauge symmetries associated with the vector multiplets, namely under $\sum_{a=1}^3 A^a$. All fields are neutral under the remaining $G_0 = U(1)^2 \subset ISO(7)$. On the other hand, $\sigma$ plays the role of a St\"uckelberg field that gives mass to the graviphoton $A^0$.

The magnetic gauging $m$ is induced by the Romans mass in the massive Type IIA uplift of this theory \cite{Guarino:2015jca}. It has the effect to mix the graviphoton $A^0$ with its magnetic dual $\wt A_0$, and in the Lagrangian it induces a topological term which requires the use of an auxiliary 2-form field $\cB^0$ \cite{deWit:2005ub}. This produces an extra Abelian 1-form gauge symmetry with parameter $\xi^0$, such that:
\be
\label{1formgauge}
\cB^0 \to \cB^0-d\xi^0 \;,\qquad A^0 \to A^0 + \frac{1}{2} m\,\xi^0 \;,\qquad \wt A_0 \to \wt A_0+\frac{1}{2}g\,\xi^0 \;.
\ee
This symmetry will be useful later when studying the BPS equations.

\subsection{Black hole horizons}

We consider static BPS black holes with dyonic charges and horizons given by a compact Riemann surface $\Sigma_\fg$. In particular we can have spherical ($S^2$, $\fg=0$), flat toroidal ($T^2$, $\fg=1$) or hyperbolic (locally $\bH^2$, $\fg>1$) horizons. The metric ansatz takes the form
\be
ds^2 = - e^{-2U(r)} dt^2 + e^{2U(r)} dr^2 + e^{2\left( \rule{0pt}{.6em} \psi(r)- U(r)\right)} ds^2_{\Sigma_\fg}
\ee
in terms of radial functions $U, \psi$. Here $ds^2_{\Sigma_\fg}$ is the metric on $\Sigma_\fg$ with constant scalar curvature $R_{\Sigma_\fg} = 2\kappa$ and $\kappa=1$ for $\fg=0$, $\kappa =0$ for $\fg=1$, $\kappa = -1$ for $\fg>1$. Locally we can take
\be
ds^2_{\Sigma_\fg} = d\theta^2 + f_\fg^2(\theta)\, d\varphi^2 \;,\qquad\qquad f_\fg(\theta) = \left\{ \begin{array}{ll} \sin\theta & \fg=0 \\
1 & \fg=1 \\
\sinh\theta & \fg>1 \;. \end{array} \right.
\ee
The scalars are taken to have radial dependence. The ansatz for the gauge fields $\cA^M$ is such that it fixes the electric charges $e_\Lambda$ and the magnetic charges $p^\Lambda$ of the black hole. In particular
\be
\label{def of charges}
p^\Lambda = \frac1{\Vol(\Sigma_\fg)} \int_{\Sigma_\fg} H^\Lambda
\;,\qquad\qquad e_\Lambda = \frac1{\Vol(\Sigma_\fg)} \int_{\Sigma_\fg} G_\Lambda \;,
\ee
where $H^\Lambda = dA^\Lambda + \delta^{\Lambda0} \frac12 m \cB^0$ and $G_\Lambda = 8\pi G_N \, \delta(\mathscr{L} \dvol_4)/\delta H^\Lambda$. The correction term ensures that the charges are gauge invariant, however it is always possible to choose a gauge in which the 2-form $\cB^0$ vanishes. The volume of $\Sigma_\fg$ is
\be
\Vol(\Sigma_\fg) = 2\pi \eta \;,\qquad\qquad \eta = \left\{ \begin{array}{ll} 2|\fg-1| & \text{for } \fg\neq 1 \\ 1 & \text{for } \fg=1 \;. \end{array} \right.
\ee
We collect the electric and magnetic charges into a symplectic vector (in general $r$ dependent)
\be
\cQ = (p^\Lambda, e_\Lambda) \;.
\ee
It will be convenient to define also
\be
\cQ^x = \langle \cP^x, \cQ \rangle \;,
\ee
which is an $SU(2)$ triplet of scalars.

To find BPS solutions we should specify an ansatz for the Killing spinors as well. The condition such that the gauge connections cancel the spin connection in the gravitini variations boils down to (see \eg{} \cite{Halmagyi:2013sla})
\be
\label{cancel spin connection}
\kappa \, \epsilon_A = - \cQ^x (\sigma^x)\du{A}{B} \Gamma^{\hat t\hat r} \epsilon_B \;.
\ee
Here $\epsilon_A$ is a doublet of spinors, $A=1,2$ is an index in the fundamental of $SU(2)$ and hatted indices correspond to vielbein. By taking the square of this equation we obtain the constraint
\be
\label{weaker BPS constraint charges}
\cQ^x \cQ^x = \kappa^2 \;.
\ee
For $\kappa\neq 0$, (\ref{cancel spin connection}) halves the number of preserved supercharges. As we will see, in the near-horizon region one finds $\cQ^\pm =0$. Using local $SU(2)$ rotations we could always enforce this condition on the whole solution. Then, in order to solve (\ref{cancel spin connection}), we would impose the projector%
\footnote{Such a projector corresponds to the one imposed by the topological twist in the boundary theory.}
\be
\label{eps projector}
\epsilon_A = (\sigma^3)\du{A}{B} \Gamma^{\hat t\hat r} \epsilon_B \;.
\ee
This gives us the stronger constraint
\be
\label{BPS constraint charges}
\cQ^3 = - \kappa \;,
\ee
which will turn out to be the BPS constraint on the charges.%
\footnote{This is equivalent to the BPS constraint $\langle \cG, \cQ\rangle = -\kappa$ in the case without hypermultiplets. In that case $\cG = (g^\Lambda, g_\Lambda)$ is the symplectic vector of magnetic and electric gaugings, also called Fayet-Iliopoulos terms.}
In practice we will not work with this rotated frame, both because we want to keep the moment maps in their simple form (\ref{moment maps}), and because in any case we will only consider near-horizon solutions. Had we chosen the opposite sign in (\ref{eps projector}), we would have considered anti-BPS solutions and the constraint (\ref{BPS constraint charges}) would have had the opposite sign. For $\kappa=0$ we are led to the same constraint (\ref{BPS constraint charges}), however it seems that there is no need to impose projectors. Nevertheless, the projector (\ref{eps projector})---or the one with opposite sign---is imposed by requiring the gaugini variations to vanish for generic charges. From a careful analysis of the BPS equations one derives another constraint \cite{Klemm:2016wng}:
\be
\label{BPS constraint vectors}
\cK^u h_{uv} \langle \cK^v, \cQ\rangle = 0 \;.
\ee
This will be useful later.

The only full black hole solution that has been constructed in this theory to date has equal magnetic and electric charges \cite{Guarino:2017eag}. However, near-horizon geometries are much easier to construct---thanks to the attractor equations \cite{DallAgata:2010ejj, Halmagyi:2013sla, Klemm:2016wng}---and they have been explicitly constructed in \cite{Guarino:2017pkw}. Since the near horizon geometry is all we need to determine the Bekenstein-Hawking entropy of the black hole, we will restrict to that. In fact, as we will see, we do not even need to find the full near-horizon geometries explicitly in order to exhibit a match with the microscopic field theory computation.

The near-horizon geometry is AdS$_2 \times \Sigma_\fg$, corresponding to the functions
\be
e^{2U} = \frac{r^2}{L^2_{\text{AdS}_2}} \;,\qquad\qquad\qquad e^{2(\psi - U)} = L^2_{\Sigma_\fg} \;,
\ee
while all scalars are constant. The full near-horizon solutions are fixed by attractor equations \cite{DallAgata:2010ejj, Halmagyi:2013sla, Klemm:2016wng}. Let us define%
\footnote{The definition of $\cL$ here differs from the more common one $\cL_\text{there} = \langle \cQ^x \cP^x, \cV \rangle$ that is used, for instance, in \cite{Klemm:2016wng}. The one here, used \eg{} in \cite{Halmagyi:2013sla, Erbin:2014hsa}, allows us to treat all cases $\kappa = \{1,0,-1\}$ uniformly.}
\be
\cZ(z^a; p^\Lambda, e_\Lambda) = \langle \cQ, \cV \rangle \;,\qquad\qquad \cL(z^a, q^u) = \langle \cP^3, \cV \rangle \;.
\ee
Then the BPS equations imply
\be
\label{BPS equation hypers}
\langle \cK^u, \cV \rangle = 0
\ee
as well as
\be
\label{attractor equations}
\partial_a\, \frac{\cZ}{\cL} = 0 \;,\qquad\qquad\qquad -i \frac{\cZ}{\cL} = L^2_{\Sigma_\fg} \;,
\ee
supplemented by the constraints (\ref{BPS constraint charges}) and (\ref{BPS constraint vectors}). In the equation above, $\partial_a$ is a derivative with respect to the vector multiplet scalars $z^a$. The first equation is in fact equivalent to $D_a \big( \cZ - i L^2_{\Sigma_\fg} \cL \big) = 0$, when combined with the second one. Moreover the second equation computes the horizon area.

Our strategy will be to use the equations to fix the hypermultiplet scalars and enforce the constraints they impose on the vector multiplet scalars and the charges, but leave the remaining freedom in the vector multiplet scalars unfixed. Let us begin with (\ref{BPS equation hypers}). The vector $\cK^\phi$ is identically zero, while the other ones give
\be
e^{-\frac\cK2} \langle\cK^\sigma,\cV\rangle = g X^0 - m \cF_0 \;,\qquad e^{-\frac\cK2} \langle\cK^\zeta,\cV\rangle = - \tilde\zeta g \sum_{a=1}^3 X^a \;,\qquad e^{-\frac\cK2} \langle\cK^{\tilde\zeta},\cV\rangle = \zeta g \sum_{a=1}^3 X^a \;.
\ee
Since $\sigma$ is a St\"uckelberg field shifted by $\bR$ gauge transformations, we can gauge fix it to zero. Together with (\ref{BPS equation hypers}) we obtain, at the horizon:%
\footnote{Here we are using that $\sum_{a=1}^3 X^a = \sum_{a=1}^3 z^a \neq 0$ since $z^a$ take values on the upper half plane. However, even relaxing this condition and allowing---in principle---specific values of $z^a$ for which (\ref{BPS equation hypers}) is solved leaving $\zeta, \tilde\zeta$ unconstrained, for $\kappa = \pm1$ we still find that (\ref{weaker BPS constraint charges}) and (\ref{BPS constraint vectors}) imply $\zeta = \tilde\zeta = 0$. We conclude that there exist no special solutions to (\ref{BPS equation hypers}) besides (\ref{hypers and constraint product z}).}
\be
\label{hypers and constraint product z}
\sigma = \zeta = \tilde\zeta = 0 \;,\qquad\qquad\qquad \prod_{a=1}^3 z^a = - \frac mg \;.
\ee
Then we consider (\ref{BPS constraint vectors}). Imposing $\zeta = \tilde\zeta =0$ the only non-vanishing components are with $\Lambda=0$, either up or down. They give a constraint on the graviphoton charges:
\be
\label{constraint graviphoton charges}
m \, e_0 - g \, p^0 = 0 \;.
\ee
Finally we impose (\ref{BPS constraint charges}). When $\zeta =\tilde\zeta=0$ only $\cP^3$ is non-vanishing, while $\cP^\pm = 0$. Using (\ref{constraint graviphoton charges}) we find $\cQ^3 = g \sum_{a=1}^3 p^a$, and thus we obtain the BPS constraint on the charges
\be
\label{BPS constraint p}
\sum_{a=1}^3 p^a = - \frac\kappa g \;.
\ee

Instead of trying to solve the remaining equations in (\ref{attractor equations}) (explicit solutions can be found in \cite{Guarino:2017pkw}), we aim to reduce them to a simpler extremization problem. We evaluate the functions $\cL$ and $\cZ$ at the horizon, imposing $\zeta = \tilde\zeta = 0$:
\bea
\label{L Z with Lagrange multiplier}
\cL &= e^{\cK/2} \bigg[ - \frac12 \, e^{2\phi} \big( gX^0 - m \cF_0 \big) + g \big( X^1 + X^2 + X^3 \big) \bigg] \\
\cZ &= e^{\cK/2} \big( e_\Lambda X^\Lambda - p^\Lambda \cF_\Lambda \big) \;.
\eea
When imposing $D_a \big( \cZ -i L^2_{\Sigma_\fg} \cL \big) = \partial_a \big[ e^{-\cK/2} \big(\cZ - i L^2_{\Sigma_\fg} \cL\big) \big] = 0$ we are supposed to vary the functions with respect to \emph{independent} scalars $z^a$. However the hypermultiplet scalar $e^{2\phi}$ plays the role of a Lagrange multiplier for the second constraint in (\ref{hypers and constraint product z}), therefore we can reduce to the problem of extremizing $-i\cZ/\cL$ with respect to \emph{constrained} scalars satisfying (\ref{hypers and constraint product z}). Imposing the constraint we find
\be
- i \frac\cZ\cL = - \frac i{g^2} \, \frac{\sum_{a=1}^3 \big( g\, e_a z^a - m \, p^a/z_a \big) }{ \sum_{a=1}^3 z^a} \qquad\qquad\text{with (\ref{hypers and constraint product z})} \;.
\ee

Although not needed here, notice that the equations $\partial_a \big[ e^{-\cK/2} \big(\cZ - i L^2_{\Sigma_\fg} \cL\big) \big] = 0$ with variations with respect to independent $z^a$, combined with the constraint (\ref{hypers and constraint product z}), fix the value of the Lagrange multiplier $e^{2\phi}$, which is the last hypermultiplet scalar we had not fixed yet.

\subsection{The Bekenstein-Hawking entropy}

The Bekenstein-Hawking entropy $S_\text{BH}$ of the black holes is given by the horizon area:
\be
S_\text{BH} = \frac{\text{Area}}{4G_N} = \frac{2\pi \eta L^2_{\Sigma_\fg}}{4G_N} \;.
\ee
The attractor equations (\ref{attractor equations}) determine the area in terms of the value of $-i \cZ/\cL$ at its critical point. We can then introduce a function
\be
\label{entropy function S}
\boxed{\quad\rule[-1.5em]{0pt}{3.5em}
\cS(z^a; p^a, e_a) = - i \frac{2\pi \eta}{4G_N}\, \frac{\cZ}{\cL} = - \frac{2\pi i}{g^2} \, \frac{\eta}{4G_N} \frac{ \sum_{a=1}^3 \big( g\, e_a z^a - m \, p^a/z_a \big) }{ \sum_{a=1}^3 z^a }
\quad}
\ee
of two complex scalars, in which the three scalars satisfy $\prod_{a=1}^3 z^a = -m/g$ and the charges satisfy $\sum_{a=1}^3 p^a = -\kappa/g$. The entropy is equal to the extremal value of this function:
\be
S_\text{BH} = \cS(\wh z^a; p^a, e_a) \qquad\text{with $\wh z^a$ such that}\qquad \partial_{z^a} \cS(z^a; p^a, e_a) \Big|_{z^a = \wh z^a} = 0 \;.
\ee

We should note that, for generic values of the charges satisfying the BPS constraint (\ref{BPS constraint p}), the critical point of $\cS$ is complex, not real. Thus, generically, there is no well-defined near-horizon geometry. A necessary condition to have a good near-horizon geometry is that $\cS(\wh z^a; p^a, e_a)$ be real positive, which imposes a further polynomial constraint on the charges. This should be interpreted as a condition to have a large smooth black hole (with finite horizon area), rather than a BPS condition on states. We also note that for every choice of charges $(p^a,e_a)$---satisfying (\ref{BPS constraint p})---it is always possible to perform a common shift of $e_a$ such that $\cS(\wh z^a; p^a, e_a)$ becomes real (not necessarily positive, though). This is a shift of the R-charge of the black hole. Such a shift does not affect the extremization problem, therefore it does not change $\wh z^a$, but it shifts $\cS$ by an imaginary amount. We conclude that (before applying quantization conditions) the domain of charges $(p^a, e_a)$ leading to large smooth BPS horizons has dimension 4.%
\footnote{There are other conditions that one should check to make sure that a good near-horizon geometry has been found, for instance that $L^2_{\text{AdS}_2}$ is positive and that $\wh z^a$ live on the upper half plane. These, however, are inequalities and therefore they do not change the dimension of the domain.}

We can describe the procedure in a slightly different way. First we fix magnetic charges that satisfy the BPS constraint (\ref{BPS constraint p}), and flavor charges $e_a - e_3$ for $a=1,2$. Then we determine the unique value of the R-charge $e_R = \frac13 \sum_{a=1}^3 e_a$ such that $\cS(\wh z^a; p^a, e_a)$ is real. In other words, for given magnetic and flavor charges, there is a unique value of the R-charge such that a large smooth black hole with those charges can possibly exist. As we will see in Section \ref{sec: comparison}, this procedure has a direct counterpart in the field theory analysis.


\section{Microscopic counting in field theory}
\label{sec: FT}

The three-dimensional quantum field theory dual to massive Type IIA on $S^6$, whose consistent truncation we have been studying in the previous section, has been identified in \cite{Guarino:2015jca}. It is an $\cN=2$ supersymmetric Chern-Simons gauge theory with gauge group $SU(N)$ and level $k$ (related to the Romans mass), coupled to three chiral multiplets $X,Y,Z$ in the adjoint representation, and with a superpotential given by
\be
W = \Tr X[Y,Z] \;.
\ee
\begin{figure}[t]
\centering
\tikzset{->-/.style = {decoration = {markings, mark=at position .4 with {\arrow[scale=2,rotate=-8]{angle 60}}, mark=at position .5 with {\arrow[scale=2,rotate=-8]{angle 60}}, mark=at position .6 with {\arrow[scale=2,rotate=-8]{angle 60}}}, postaction={decorate}}}
\begin{tikzpicture}
\draw [thick] (0,0) circle [radius=.8cm] node(x) {\footnotesize$SU(N)_k$};
\draw [->-,thick] (1.2,0) ++(-141: 1.2) arc (-141: 141: 1.2);
\end{tikzpicture} \hspace{2cm} \raisebox{1.2cm}{$W = \Tr X[Y,Z]$}
\caption{Quiver diagram and superpotential of the 3d dual to massive Type IIA on $S^6$.
\label{fig: quiver}}
\end{figure}
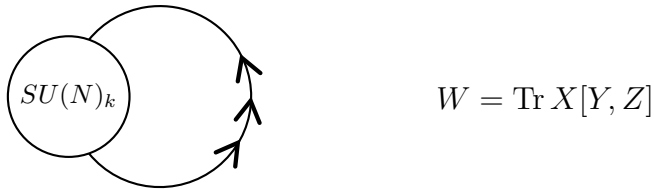
The corresponding quiver diagram is represented in Figure \ref{fig: quiver} (it coincided with the quiver diagram of 4d $\cN=4$ SYM). The global symmetry of the theory is $SU(3) \times U(1)_R$, where the latter is the R-symmetry. We can choose the following basis for the maximal torus of the global symmetry:
\be
\begin{array}{c|ccc}
& \wt{U(1)}_1 & \wt{U(1)}_2 & \wt{U(1)}_3 \\
\hline
X & 1 & 0 & 0 \\
Y & 0 & 1 & 0 \\
Z & -1 & -1 & 2
\end{array}
\ee
Here $\wt{U(1)}_{1,2}$ are flavor symmetries, and we let $J_a$ be the corresponding currents, while $\wt{U(1)}_3$ is an R-symmetry. We have chosen an R-symmetry generator that does not commute with $SU(3)$ because it gives integer charges to all fields.

The regime in which the bulk gravitational theory is weakly coupled corresponds to the large $N$ limit with $k$ fixed (or at least $N\gg k$). The BPS dyonic black hole solutions in AdS$_4$ induce, via the rules of AdS/CFT \cite{Maldacena:1997re, Gubser:1998bc, Witten:1998qj}, relevant deformations of the boundary theory. First of all the 3d CS theory is placed on $\Sigma_\fg \times \bR$, where $\Sigma_\fg$ is a Riemann surface with the same genus $\fg$ as the black hole horizon. Moreover, the theory is topologically twisted on $\Sigma_\fg$ \cite{Maldacena:2000mw} in such a way that one complex supercharge is preserved. In other words, there is a background gauge field $V$ on $\Sigma_\fg$, coupled to an R-symmetry, equal and opposite to the spin connection and therefore such that%
\footnote{We can turn on a background flux because all gauge-invariant operators have integer R-charge.}
$\frac1{2\pi} \int_{\Sigma_\fg} dV = \fg-1$. In the presence of flavor symmetries there are multiple choices one can make for the R-symmetry used in the twist. We can parametrize those choice by keeping the R-symmetry fixed, and introducing Abelian background gauge fields $F_a$ coupled to the flavor symmetry currents $J_a$. We then turn on a background (in the Cartan subalgebra, without loss of generality) for all of them:
\be
\fn_a = \frac1{2\pi} \int_{\Sigma_\fg} F_a \in \Gamma_\text{flav} \;.
\ee
The numbers $\fn_a$ are GNO quantized \cite{Goddard:1976qe} in the coroot lattice $\Gamma_\text{flav}$ of the flavor symmetry, and effectively parametrize the twist. It turns out to be convenient to introduce an auxiliary flux parameter, formally associated to the R-symmetry, that is defined linearly in terms of the other ones. Then the numbers $\fn_a$ in field theory correspond to the magnetic charges of the black hole (the precise normalization will be fixed in Section \ref{sec: comparison}). In our case we introduce $\fn_3$, besides $\fn_{1,2}$, such that $\sum\nolimits_{a=1}^3 \fn_a = 2(\fg-1)$. This description is convenient because the Weyl group of $SU(3)$ acts as permutations of the indices $a=1,2,3$.

The black hole microstates correspond to ground states of this system, therefore in order to give a microscopic account of the black hole entropy we should count those ground states \cite{Benini:2015eyy, Benini:2016rke}. This is a non-trivial problem because the theory is strongly coupled in the IR. However we can have a good estimate, in the large $N$ limit, of the number of ground states by computing an index:
\be
\label{index general}
Z(\fn_a, \Delta_a) = \Tr \, (-1)^F \, e^{-\beta H}\, e^{i \fq_a \Delta_a} \;,
\ee
where $F$ is the fermion number, $H$ is the Hamiltonian on $\Sigma_\fg$ and in the presence of magnetic flavor fluxes $\fn_a$ on $\Sigma_\fg$, $\fq_a$ are the flavor charges (a Cartan basis thereof) while $\Delta_a$ are chemical potentials. This object is a Witten index \cite{Witten:1982df}: it only receives contributions from ground states $H=0$, and it is protected by supersymmetry. It turns out that this object can be computed exactly with localization techniques \cite{Benini:2015noa, Benini:2016hjo, Closset:2016arn} (see also \cite{Nekrasov:2014xaa, Gukov:2015sna}). The index takes the form
\be
Z(\fn,\Delta) = \frac1{|\text{Weyl}|} \sum_{\fm \,\in\, \Gamma_\text{gauge}} \oint_\text{JK} Z_\text{int}(x,\fm; y,\fn) \;.
\ee
Here $|\text{Weyl}|$ is the order of the gauge Weyl group, $\Gamma_\text{gauge}$ is the co-root lattice of the gauge group, and the sum is over gauge fluxes $\fm$ on $\Sigma_\fg$. Then $Z_\text{int}$ is a meromorphic $r$-form (where $r$ is the rank of the gauge group) on the space of complexified flat gauge connections on $S^1$, which can be parametrized by gauge fugacities $x$. Finally $y_a=e^{i\Delta_a}$ are fugacities for the flavor symmetries. The integral is a contour integral along a particular contour called the Jeffrey-Kirwan residue \cite{JeffreyKirwan}. We refer to \cite{Benini:2015noa, Benini:2016hjo, Closset:2016arn} for details.

To extract the Bekenstein-Hawking entropy of the black holes, we should compute the large $N$ limit of this expression. Then, since the index (\ref{index general}) is in the grand canonical ensemble with respect to the electric charges, in order to extract the contribution from a particular charge sector we should perform a Fourier transform:
\be
\label{Fourier transform}
Z(\fn_a, \fq_a) = \int \frac{d^d \Delta_a}{(2\pi)^d} \, Z(\fn, \Delta) \, e^{-i \fq_a \Delta_a}
\ee
where $d$ is the total rank of the flavor symmetry. Here a complication arises \cite{Benini:2016rke}: since the index $Z(\fn,\Delta)$ depends on flavor fugacities but it cannot have a fugacity for the R-symmetry (it would spoil supersymmetry), what we have on the left-hand-side is the sum of contributions from all states with fixed flavor charges but arbitrary R-charge. However in the large $N$ limit we can assume that one R-charge sector will dominate. Moreover, assuming that at large $N$ the integral (\ref{Fourier transform}) can be computed in the saddle-point approximation, one defines the function
\be
\cI(\Delta_a; \fn_a, \fq_a) = \log Z(\fn_a, \Delta_a) - i \sum\nolimits_a \fq_a \Delta_a
\ee
and the logarithm of the number of states is given by
\be
\label{extremization general}
\log(\#\text{ states})(\fn_a,\fq_a) = \cI \big( \wh\Delta_a; \fn_a, \fq_a \big) \qquad\text{with $\wh\Delta_a$ such that}\quad \parfrac{\cI}{\Delta_a} \Big|_{\Delta_a = \wh\Delta_a} = 0
\ee
which is the Legendre transform of $\log Z$.

Interestingly, it has been noticed in \cite{Benini:2016rke} in one example that if we introduce one auxiliary chemical potential $\Delta_{d+1}$, formally associated to the R-symmetry, defined in terms of the other ones---as we did for the flavor fluxes $\fn_a$---such that $\sum_a \Delta_a = 2\pi$, and we change the definition of the charges such that $\fq_a - \fq_{d+1}$ are the flavor charges, then we can extract the dominant R-charge from (\ref{extremization general}) by requiring that $\cI$ be real. As R-charge we can take, for instance, $\fq_R = \frac{1}{d+1} \sum_a \fq_a$. The same will be true in the example considered here.

Another subtlety to keep in mind is that $Z$ in (\ref{index general}) is an index and thus it counts states with sign $(-1)^F$. It has been argued in \cite{Benini:2015eyy}, using the $\fsu(1,1|1)$ superconformal algebra of the AdS$_2 \times \Sigma_\fg$ near-horizon region to the black hole, that the states associated to the pure single-center BPS black holes%
\footnote{By ``pure single-center black hole'' we mean the near-horizon AdS$_2 \times \Sigma_\fg$ solution with boundary conditions that fix the microcanonical ensemble with respect to both magnetic and electric charges \cite{Sen:2008vm}.}
have $(-1)^F=1$, and thus they are precisely counted by the index. This argument is essentially the same as the one given in \cite{Sen:2009vz} (and nicely summarized \eg{} in \cite{Sen:2014aja}) for BPS black holes in flat space.
This is not true for the states coming from multi-center black holes and hair, whose number however we might expect to be subleading.

\subsection{The topologically twisted index}

Let us go back to the specific $SU(N)_k$ theory we are interested in. We use $\wt{U(1)}_{1,2}$ as a maximal torus of the flavor symmetry. We denote by $y_{1,2}$ the associated fugacities, with
\be
y_a = e^{i\Delta_a} \;,
\ee
and by $\fn_{1,2}$ the associated fluxes on $\Sigma_\fg$. In order to restore the symmetry under the Weyl group of $SU(3)$, it is convenient to introduce also the auxiliary variables $y_3$ and $\fn_3$ fixed by
\be
\label{constraints n and y}
\sum\nolimits_{a=1}^3 \fn_a = 2(\fg-1) \;,\qquad\qquad\qquad \prod\nolimits_{a=1}^3 y_a = 1 \;.
\ee
In order to avoid the technicality arising from the structure of the Cartan subalgebra of $\fsu(N)$, we consider the theory with gauge group $U(N)_k$ instead. The computation of the $U(N)$ partition function is simpler, and in our case it provides the same result as the $SU(N)$ theory. In fact it has been proven in \cite{Benini:2015noa} that the index of a $U(N)_k$ CS theory with no topological flux is exactly equal to the index of the corresponding $SU(N)_k$ CS theory whenever the matter is neutral under the central $U(1)$ in $U(N)$. Following the rules in \cite{Benini:2015noa, Benini:2016hjo, Closset:2016arn} and after some manipulations, the index takes the form
\begin{multline}
Z(\fn,\Delta) = \frac{(-1)^N}{N!} \prod_{a=1}^3 \frac{y_a^{N^2(1+\fn_a-\fg)/2}}{(1-y_a)^{N(1+\fn_a-\fg)}} \sum_{\fm\in\bZ^N} \oint_\text{JK} \prod_{i=1}^N \frac{dx_i}{2\pi i x_i} \, x_i^{k\fm_i} \times \\
\times \prod_{j(\neq i)}^N \prod_{a=1}^3 \Big( \frac{x_i - y_a x_j}{x_j - y_a x_i} \Big)^{\fm_i} \prod_{i\neq j}^N \Big( 1 - \frac{x_i}{x_j} \Big)^{1-\fg} \prod_{a=1}^3 \Big( 1 - y_a \frac{x_i}{x_j} \Big)^{\fg - 1-\fn_a} \;.
\end{multline}
The integrand has poles at $x_i = 0$ and $\infty$ (for generic values of $y_a$). Assuming $k>0$, the JK prescription selects an integration contour around $x_i=0$ and thus the integral computes minus the sum of the residues there. Since there are poles at $x_i=0$ only for $\fm_i \leq M-1$ for some large positive $M$, we can restrict the sum to those values and resum the geometric series before picking the residues. This leads to the expression
\be
\label{TT index}
Z = \frac1{N!} \prod_{a=1}^3 \frac{y_a^{N^2(1+\fn_a-\fg)/2}}{(1-y_a)^{N(1+\fn_a-\fg)}} \sum_{I \,\in\, \text{BAE}} \big( \det\bB)^{\fg-1} \prod_{i\neq j}^N \Big( 1 - \frac{x_i}{x_j} \Big)^{1-\fg} \prod_{a=1}^3 \Big( 1- y_a\, \frac{x_i}{x_j} \Big)^{\fg-1-\fn_a} \;.
\ee
Here $I$ runs over the solutions to the ``Bethe Ansatz Equations'' (BAEs)
\be
\label{BAEs}
1 = e^{iB_i(x)} = x_i^k \prod_{j(\neq i)}^N \prod_{a=1}^3 \frac{x_i - y_a x_j}{x_j - y_a x_i} \qquad\qquad\forall i = 1,\dots, N \;,
\ee
while the Jacobian matrix $\bB$ is given by
\be
\bB_{ij} = \frac{\partial \, e^{iB_i(x)}}{\partial\log x_j} \;.
\ee
This matrix can be written in a more explicit form as
\be
\bB_{ij} = e^{iB_i(x)} \bigg[ \bigg( k + \sum_{l=1}^N D_{il} \bigg) \delta_{ij} - D_{ij} \bigg] \;, \quad D_{ij} = z \parfrac{}{z} \log \Big( \frac{z-y_1}{1-y_1 z}\, \frac{z-y_2}{1-y_2 z}\, \frac{z-y_3}{1-y_3z}\Big) \bigg|_{z = \frac{x_i}{x_j}} \;.
\ee
We stress that (\ref{TT index}) is an exact expression for the index, valid at finite $N$.

The BAEs (\ref{BAEs}) are $N$ algebraic equations in $N$ complex variables $x_i$: in general they have a large number of solutions and cannot be analytically solved. However for any solution $\{x_i\}$ we can generate other ones $\{\omega x_i \}$ where $\omega$ is a $k$-th root of unity. Each of the $k$ solutions in the orbit gives the same contribution to (\ref{TT index}).

It is convenient to perform the change of variables
\be
x_i = e^{iu_i} \;,\qquad\qquad y_a = e^{i\Delta_a} \;,
\ee
where $\Delta_a$ are chemical potentials for the flavor symmetries. The angular variables are defined modulo $2\pi$, and the constraint on $y_a$ becomes $\sum_{a=1}^3 \Delta_a \in 2\pi \bZ$. The BAEs in the new variables take the form
\be
ku_i + i \sum_{j=1}^N \sum_{a=1}^3 \Big[ \Li_1\big( e^{i(u_j-u_i+\Delta_a)} - \Li_1 \big( e^{i(u_j-u_i-\Delta_a)} \big) \Big] - 2\pi n_i + \pi N = 0 \;,
\ee
where the integers $n_i$ express the angular ambiguity, while
\be
\Li_s(z) = \sum_{k=1}^\infty \frac{z^k}{k^s}
\ee
are the polylogarithm functions and $\Li_1(z) = -\log(1-z)$. The BAEs can then be obtained as the critical point equations of a function $V_\text{B}$ that we call the ``Bethe potential'', or equivalently ``Yang-Yang functional'' \cite{Yang:1968rm}:
\be
\label{Bethe potential}
V_\text{B} = - \sum_{i=1}^N \frac k2 u_i^2 + \frac12 \sum_{i,j=1}^N \sum_{a=1}^3 \Big[ \Li_2\big( e^{i(u_j - u_i + \Delta_a)} \big) - \Li_2\big( e^{i(u_j - u_i - \Delta_a)} \big) \Big] + \sum_{i=1}^N \pi m_i u_i \;,
\ee
where now the integers $m_i$ incorporate the various angular ambiguities.

\subsection[The large $N$ limit]{The large \matht{N} limit}

We proceed by computing (\ref{TT index}) in the large $N$ limit at fixed $k$. The computation is essentially the same as the one in \cite{Benini:2015eyy}, and turns out to be very similar to the computation of the large $N$ limit of the $S^3$ partition function in \cite{Herzog:2010hf, Jafferis:2011zi}. More examples have been considered in \cite{Hosseini:2016ume} and a rather general analysis have been performed in \cite{Hosseini:2016tor}, therefore here we will be brief.

First of all, we assume that there is one $k$-fold orbit of solutions to the BAEs that dominates $Z$. To determine it, we consider a continuous distribution of points $u(t)$, where $t$ is the continuous version of the discrete index $i=1,\dots, N$, and a density distribution $\rho(t)$ defined by
\be
\rho(t) = \frac1N\, \frac{di}{dt} \;.
\ee
In the continuum approximation, sums over $i$ are turned into integrals: $\sum_{i=1}^N \to N \int dt\, \rho(t)$, and the density distribution is normalized as $\int dt\, \rho(t) = 1$. From numerical solutions to (\ref{BAEs}), and as suggested by \cite{Herzog:2010hf, Jafferis:2011zi}, we consider the following ansatz for the behavior of the dominant solution:
\be
u(t) = N^\alpha \big( it + v(t) \big) \;,
\ee
where $v(t)$ is real and $\alpha$ is an exponent to be determined. Then we compute the large $N$ limit of the Bethe potential $V_\text{B}$, as a functional of $u(t)$ and $\rho(t)$. On general grounds, we know that the index $Z$ is analytic in $\Delta_a$ \cite{Benini:2015noa}, therefore it is convenient to perform all computations with $\Delta_a \in \bR$ and analytically continue the result at the end. Only for a specific set of values of the integers $m_i$ there is a cancelation of ``long range forces'' in (\ref{Bethe potential}) and the large $N$ functional becomes local:
\be
\pi m_i = \bigg( \sum\nolimits_a \Delta_a - 3\pi \bigg) \sum_{j=1}^N \bigg[ \Theta \big( \im(u_i - u_j) \big) - \Theta\big( \im(u_j - u_i) \big) \bigg] \;.
\ee
Here $\Theta$ is the Heaviside function.

The functional $V_\text{B}(r,\rho; \mu)$ is, at leading order in $N$:
\begin{multline}
\label{Bethe potential large N}
V_\text{B}(v,\rho; \mu) = N^{1+2\alpha} \int dt \bigg[ -ik\, t\, \rho(t)\, v(t) - \frac k2 \rho(t) \big( v(t)^2 - t^2 \big) \bigg] \\
+ N^{2-\alpha} \bigg[ i G(\Delta) \int dt\, \frac{\rho(t)^2}{1-i\, \dot v(t)} - i \mu \Big( \int dt\, \rho(t) - 1 \Big) \bigg] \;.
\end{multline}
The function $G(\Delta)$ is defined as%
\footnote{Such a function appears because $\Li_2(e^{iu}) + \Li_2(e^{-iu}) = g'_+(u)$ for $0 < \re u < 2\pi$.}
\be
G(\Delta) = \sum_{a=1}^3 g_+(\Delta_a) \;,\qquad\qquad g_+(x) = \frac{x^3}6 - \frac\pi2 x^2 + \frac{\pi^2}3 x \;.
\ee
We have enforced the normalization condition $\int \rho = 1$ with a Lagrange multiplier $\mu$, and we have chosen its scaling with $N$ for convenience. The dominant solution to the BAEs in the large $N$ limit is obtained by extremizing $V_\text{B}$. Only for $\alpha = \frac13$ there is a competition between the various terms and a well-behaved saddle point is found.%
\footnote{Moreover, in this scaling argument we are assuming that $k$ does not scale with $N$.}
In this case the Bethe potential scales as $N^{5/3}$.

The BAEs correspond to the system $\delta V_\text{B}/\delta v(t) = \delta V_\text{B}/\delta \rho(t) = 0$, together with the normalization condition $\partial V_\text{B}/\partial\mu=0$. After some manipulations, the first two equations reduce to
\be
\label{extremization V_B equation}
\mu = - k\, t\, v(t) + \frac{ik}2 \big( v(t)^2 - t^2 \big) + \frac{2G(\Delta) \, \rho(t)}{1 - i \, \dot v(t)} \;.
\ee
We solve this equation taking $k>0$ as well as
\be
\label{constraint Deltas}
0< \Delta_a < 2\pi \qquad\text{and}\qquad \sum\nolimits_{a=1}^3 \Delta_a = 2\pi \;,
\ee
which implies also $G(\Delta)>0$. We look for solutions in which $\rho(t)$ is positive, bounded, and either integrable or with compact support between two zeros. It turns out that there exists only one solution satisfying these requirements, and it has compact support. After fixing the normalization $\int \rho=1$, the solution is
\bea
\label{BAE solution large N}
v(t) &= - \frac t{\sqrt 3} & \mu &= \frac{3^{7/6} k^{1/3} G(\Delta)^{2/3}}4 \big( 1 - i /\sqrt 3\big) \\
\rho(t) &= \frac{3^{1/6} k^{1/3}}{2G(\Delta)^{1/3}} - \frac{2kt^2}{3\sqrt3\, G(\Delta)} \qquad\qquad & t_\pm &= \pm \frac{3^{5/6} G(\Delta)^{1/3}}{2k^{1/3}}
\eea
with support on the interval $\cD=[t_-, t_+]$. The density $\rho(t)$ vanishes at $t_\pm$.

We notice that the $k$-fold degeneracy of the solutions is invisible in the large $N$ limit: the $k$ solutions in the orbit are related by shifts of $v(t)$ by $2\pi / kN^{1/3}$. The solution for $\sum_{a=1}^3 \Delta_a = 4\pi$ is similar to (\ref{BAE solution large N}): just map $v(t) \to -v(t)$, $G(\Delta) \to -G(\Delta)$ and $\mu \to\mu^*$. The density $\rho(t)$ is well-defined because $G(\Delta)<0$ in this range. The cases $\sum_a \Delta_a = 0, 6\pi$ imply $\Delta_a = 0, 2\pi$ respectively (since $0\leq \Delta_a \leq 2\pi$) and do not lead to solutions to the BAEs.

It was proven in \cite{Hosseini:2016tor} that, for a large class of quiver gauge theories including the one we are studying here, the following relation holds:
\be
V_\text{B} (\Delta_a) \Big|_\text{BAEs} = i \,\frac35\, N^{5/3} \mu(\Delta_a) \;.
\ee
On the left-hand-side is the Bethe potential (\ref{Bethe potential large N}) evaluated on the solution (\ref{BAE solution large N}). The relation is indeed satisfied in our case. If we restrict to $\sum_a \Delta_a = 2\pi$ there is also a connection with the $S^3$ partition function $F_{S^3}$ \cite{Kapustin:2009kz, Jafferis:2010un, Hama:2010av} of the gauge theory, in the large $N$ limit:
\be
V_\text{B}(\Delta_a) \Big|_\text{BAEs} = \frac{i\pi}2\, F_{S^3} \big( R_a = \Delta_a/\pi \big) \;,
\ee
where $R_a$ are the R-charges. The $S^3$ partition function of the gauge theory we are studying here has been considered \eg{} in \cite{Fluder:2015eoa}.

The last step is to compute the large $N$ limit of the expression (\ref{TT index}) for $Z$, as a functional of the solutions $(v,\rho)$ to the BAEs, and then to plug in the dominant solution (\ref{BAE solution large N}) we found. Once again, the computation is essentially as the one in \cite{Benini:2015eyy}.%
\footnote{In \cite{Benini:2015eyy} it was crucial to keep into account ``exponential tails''. In the cases where $\log Z$ scales like $N^{5/3}$ such tails do not play a role \cite{Hosseini:2016tor}.}
It turns out that at large $N$ the logarithm of the index grows as $N^{5/3}$. In particular the $k$-fold degeneracy of the solutions is irrelevant at leading order in $N$. As a functional of the solutions to the BAEs and at leading order in $N$, the index is given by:
\be
\log Z(\fn, \Delta; v, \rho) = - N^{5/3} f_+(\fn, \Delta, \fg) \int_\cD dt\, \frac{\rho(t)^2}{1-i\, \dot v(t)}
\ee
with
\be
f_+(\fn,\Delta, \fg) = \sum_{a=1}^3 \big( \fg - 1 - \fn_a \big)\, g_+'(\Delta_a) + \frac{(1-\fg)\pi^2}3 \;.
\ee
Plugging in the solution (\ref{BAE solution large N}) we find
\be
\log Z(\fn, \Delta) = - \frac{3^{7/6}}{10}\, \frac{f_+(\fn, \Delta,\fg)}{G(\Delta)^{1/3}}\, k^{1/3} N^{5/3} \big( 1 - i/\sqrt 3 \big) \;.
\ee
This expression can be further simplified recalling that $\fn_a$ and $\Delta_a$ are constrained by (\ref{constraints n and y}). Specializing to the case in which%
\footnote{The solution for the other case, in which $\sum_a \Delta_a = 4\pi$, can be obtained from this one simply mapping $\Delta_a \to 2\pi - \Delta_a$.}
$\sum\nolimits_{a=1}^3 \Delta_a = 2\pi$, one finds that
\be
f_+(\fn, \Delta,\fg) = - \frac12 \Delta_1 \Delta_2 \Delta_3 \sum\nolimits_{a=1}^3 \frac{\fn_a}{\Delta_a} \;,\qquad\qquad G(\Delta) = \frac12 \Delta_1 \Delta_2 \Delta_3 \;.
\ee
We are thus led to the simple expression:
\be
\label{log Z large N final}
\boxed{\quad\rule[-1.6em]{0pt}{3.7em}
\log Z(\fn, \Delta) = \frac{3^{7/6}}{2^{5/3} 5} \big(1-i/\sqrt3\big) \; k^{1/3} N^{5/3} \big( \Delta_1 \Delta_2 \Delta_3\big)^{2/3} \sum_{a=1}^3 \frac{\fn_a}{\Delta_a} \;.
\quad}\ee
This expression seems not to depend on $\fg$, however recall that the fluxes $\fn_a$ are constrained as in (\ref{constraints n and y}) and that introduces the dependence on $\fg$.

In fact, the general analysis of \cite{Hosseini:2016tor} gives a compact way to compute the index once the dominant solution to the BAEs is found:
\be
\log Z = \frac35 N^{5/3} \sum_a \fn_a \, \parfrac{\mu(\Delta)}{\Delta_a} \;.
\ee
The expression (\ref{log Z large N final}) agrees with this one.


\section{Entropy matching through attractor equations}
\label{sec: comparison}

We compare the Bekenstein-Hawking entropy computed from supergravity in Section \ref{sec: SUGRA} with the microstate counting from field theory in Section \ref{sec: FT}.

First of all we need a dictionary between the charges. In field theory there are three electric and magnetic charges $(\fn_a, \fq_a)$ that are integer, and $\fn_a$ satisfy the BPS constraint (\ref{constraints n and y}). To understand the quantization condition of $(p^a, e_a)$ in supergravity (see a similar discussion in \cite{Benini:2016rke}) we recall that the Yang-Mills action is normalized in the same way as the Einstein-Hilbert term. Rescaling to canonical normalization we find
\be
\fn_a = \eta g\, p^a \,\in \bZ \;,\qquad\qquad \fq_a = \frac{\eta}{4G_N g} \, e_a \,\in \bZ \;.
\ee
This is compatible with (\ref{BPS constraint p}).

Then we need a dictionary between the field theory chemical potentials $\Delta_a$, constrained by (\ref{constraint Deltas}), and the supergravity vector multiplet scalars $z^a$, constrained by (\ref{hypers and constraint product z}). We propose
\be
\label{map Delta z}
\Delta_a = \frac{2\pi z^a}{\sum_{a=1}^3 z^a} \;.
\ee
This automatically guarantees $\sum_a \Delta_a = 2\pi$. The map (\ref{map Delta z}) is three-to-one (before taking into account that the scalars $z^a$ take values in the upper half-plane), not invertible: a common rotation of $z^a$ by $e^{2\pi i/3}$ leaves the $\Delta_a$'s invariant. This resonates with the fact that the large $N$ index (\ref{log Z large N final}) is not a single-valued function in the complex $\Delta_a$-plane. The inverse of (\ref{map Delta z}) is
\be
z^a = e^\frac{i\pi}{3} \Big( \frac mg \Big)^{1/3} \, \frac{\Delta_a}{(\Delta_1 \Delta_2 \Delta_3)^{1/3}}
\ee
which has in fact three sheets%
\footnote{It is important that one uses the same branch of the root for the three values $a=1,2,3$.}
and automatically guarantees $\prod_{a=1}^3 z_a = -m/g$. One also obtains the relation $\big( \prod_a \Delta_a \big)^{1/3} \big( \sum_a z_a \big) = e^{i\pi/3} \, 2\pi \, (m/g)^{1/3}$.

Finally we need a dictionary between the field theory dimensionless parameters $N$ and $k$ and the supergravity dimensionful parameters $g$, $m$ and $G_N$ \cite{Guarino:2015jca}:
\be
\frac{m^{1/3} g^{-7/3}}{4G_N} = \frac{3^{2/3}}{2^{2/3}\, 5} \, k^{1/3} N^{5/3} \;,\qquad\qquad \frac{16\pi^3}3 \Big( \frac{m}{g} \Big)^5 = N k^5 \;.
\ee
Although not needed here, the relation with the Type IIA mass parameter is $k = 2\pi \ell_s m$.

Consider now the index function $\cI(\Delta_a; \fn_a, \fq_a) = \log Z(\fn, \Delta) - i \fq_a \Delta_a$ whose value at the critical point computes the large $N$ ground state degeneracy:
\be
\cI = \frac{3^{2/3} \, e^{-i\pi/6}}{2^{2/3} \, 5} \, k^{1/3} N^{5/3} \, \big( \Delta_1 \Delta_2 \Delta_3 \big)^{2/3} \sum_{a=1}^3 \frac{\fn_a}{\Delta_a} - i \sum_{a=1}^3 \fq_a \Delta_a \;.
\ee
Using the dictionaries for the various quantities we can rewrite it as
\be
\cI = \cS = - \frac{2\pi i}{g^2} \, \frac{\eta}{4G_N} \frac{ \sum_{a=1}^3 \big( g\, e_a z^a - m \, p^a/z_a \big) }{ \sum_{a=1}^3 z^a } \;,
\ee
exactly matching the entropy function $\cS$ in (\ref{entropy function S}) we found in supergravity. Notice in particular that the supergravity variables $z^a$ provide a global description of the parameter space, on which the function $\cI=\cS$ is single valued.

Summarizing, we have reduced the classical supergravity computation of the horizon area and the quantum field theory computation of the ground state degeneracy---more precisely, of its index---to the same extremization problem: finding the value of a complex function at its critical point. Since the two functions $\cS$ and $\cI$ coincide (as functions of variables with the same constraint), the result is guaranteed to be the same: the black hole entropy exactly equals the ground state degeneracy at leading order.

Notice that from the field theory index we can also reproduce the R-charge of the black holes, along the lines of \cite{Benini:2016rke}. In field theory the flavor charges are $\fq_a - \fq_3$. Keeping them fixed, we perform a common shift of the $\fq_a$'s  (which does not affect the extremization problem since $\sum_{a=1}^3\Delta_a=2\pi$) in such a way that the value of $\cI$ at the critical point becomes real. Then we read off the R-charge $\fq_R = \frac13 \sum_a \fq_a$. Exactly the same procedure fixes the black hole R-charge $e_R = \frac13 \sum_{a=1}^3 e_a$ in supergravity, as we commented upon at the end of Section \ref{sec: SUGRA}.


\section{Conclusions}
\label{sec: conclusions}

In this paper we have studied the entropy of static dyonic BPS black holes in 4d $\cN=2$ gauged supergravities with vector and hyper multiplets. We have focused on a specific example: BPS black holes in AdS$_4\times S^6$ in massive Type IIA. We have shown that, similarly to the case with no hypermultiplets, the entropy can be expressed as the value of a function $\cS$ at its critical point. Moreover we have shown that the entropy can be reproduced with a microscopic computation in the dual (via AdS/CFT) 3d QFT: there the logarithm of the number of states can be reduced to the very same extremization problem.

It would be interesting to understand the case with hypermultiplets more in general. The hypermultiplets can give mass to some of the vector multiplets, thus effectively reducing the extremization problem to a submanifold of the vector multiplet scalar manifold $\cM_\text{SK}$. Only this submanifold seems to be visible to the QFT index. Presumably, a general matching argument (similar to the one presented in the Introduction for the cases with no hypermultiplets) would involve not only the prepotential on $\cM_\text{SK}$, but also the Killing vector fields on the hypermultiplet scalar manifold $\cM_\text{QK}$ that are gauged, and the embedding tensor. How these quantities appear on the QFT side is unclear to us.


\section*{Acknowledgments} 

We thank Alberto Zaffaroni for many discussions on related topics. F.B. is supported in part by the MIUR-SIR grant RBSI1471GJ ``Quantum Field Theories at Strong Coupling: Exact Computations and Applications'', and by the IBM Einstein Fellowship at the Institute for Advanced Study. H.K. and P.M. thank the Institute for Advanced Study for hospitality and support during the early stages of this work.


\bibliographystyle{ytphys}
\bibliography{BHentropy}
\end{document}